\documentclass[12pt,preprint]{aastex}

\shortauthors{Wang et al.}
\begin{document}

\title{On the fraction of X-ray obscured quasars in the local universe}
\author{J. X. Wang \& P. Jiang}
\affil{Center for Astrophysics, University of Science and Technology of China, Hefei, Anhui 230026, P. R. China \\
Joint Institute of Galaxies and Cosmology, USTC and SHAO, CAS
}
                                                                                
\email{jxw@ustc.edu.cn}

\begin{abstract}
Recent wide area hard X-ray and soft Gamma ray surveys have shown that
the fraction of X-ray obscured AGNs in the local universe significantly
decreases with intrinsic luminosity. 
In this letter we point out that two correction have to be made to the
samples: 
1) radio loud AGNs have to be excluded since their X-ray
emission might be dominated by the jet component;
2) Compton thick sources have to be excluded too since their
hard X-ray and soft gamma ray emission are also strongly attenuated by Compton scattering.
The soft gamma-ray selected AGN samples obtained by $SWIFT$ and 
$INTEGRAL$ provide the best opportunity to study the fraction of obscured 
AGN in the local universe in the least biased way. 
We choose these samples to check if the corrections
could alter the above result on the fraction of obscured AGNs.
We find that before the corrections both samples show significant 
anti-correlation between $L_X$ and $N_H$, indicating obvious decrease in
the fraction of obscured AGNs with 
luminosity. However, after the corrections, we find only marginal evidence
of anti-correlation (at 98\% confidence level) in the SWIFT sample, and no
evidence at all in the $INTEGRAL$ sample which consists of comparable number
of objects. We conclude that current samples only show a marginal decrease
in the fraction of obscured AGNs in the local universe,
and much larger samples are required to
reach a more robust conclusion.
\end{abstract}

\section{Introduction}
A large population of obscured powerful quasars called type-2 quasars has
long been predicted by the widely accepted unified model for active galactic 
nuclei (AGNs; Antonucci 1993).
They are believed to be intrinsically the same as type-1 
quasars but with strong obscuration in both optical and X-ray band due to
the presumable torus.
Most of such type-2 quasars, which might dominate the black hole growth
(e.g., Mart\'\i nez-Sansigre et al. 2005), have been missed by optical 
surveys for quasars. Doubt has been expressed at times on the existence of
type-2 quasars (e.g. Halpern, Turner \& George 1999), mainly due to their 
rareness and the fact that they are easily mimicked by other types of AGNs.
The hard X-ray emission is less affected by the photon-electric absorption, 
making hard X-ray
surveys good approach to search for type-2 quasars. 
Another advantage of X-day data is that X-ray spectral fitting can yield
intrinsic luminosity and absorption simultaneously, which are both
required to study the population of obscured quasars.
A number of obscured quasars
have been successfully revealed by recent deep and wide-area X-ray surveys 
performed by $Chandra$ and XMM 
(Norman et al. 2002; Stern et al. 2002; 
Mainieri et al. 2002; Fiore et al. 2003; Caccianiga et al. 2004).

However, many studies have claimed that the fraction of obscured quasars is 
much smaller than expected from the simplified AGN unified model. 
Ueda et al. (2003) computed the X-ray luminosity 
function for 2 -- 10 keV selected AGN samples, and found that the fraction of
X-ray absorbed AGNs (with $N_H$ $>$ 10$^{22}$ cm$^{-2}$) drops from $\sim$
0.6 at intrinsic $L_X$ around 10$^{42}$ ergs s$^{-1}$ to around 0.3
at $L_X$ above 10$^{44}$ ergs s$^{-1}$.
Several other studies based on hard X-ray surveys also
support the scheme that the fraction of type II AGNs (or X-ray absorbed
AGNs) decreases with intrinsic luminosity (Steffen et al. 2003; Barger et al. 
2005; La Franca et al. 2005; Shinozaki et al. 2006).
However, contrary results are also reported. 
Perola et al. (2004) found that
the fraction of obscured AGNs in HELLAS2XMM does not change with X-ray 
luminosity, although the fraction (40\%) appears smaller than expected.
Eckart et al. (2006) found that half of the AGNs identified by the
SEXSI program are X-ray obscured with $N_H$ $>$ 10$^{22}$ cm$^{-2}$,
and the fraction of obscured AGNs is independent of the unobscured luminosity.
The discrepancy can be attributed to the redshift evolution in the
fraction of obscured quasars (La Franca et al. 2005; Wang et al. 2006):
while shallow and wide surveys are probing quasars in the local universe
where probably only a small fraction of quasars are obscured,
deep surveys are detecting quasars at much higher redshift where
more than half of them are obscured (Wang et al. 2006).

Soft gamma-ray selected AGN samples (Markwardt et al. 2005, hereafter M06; Bassani et al.  2006, hereafter B06; Beckmann, Gehrels \& Shrader 2006) obtained by $SWIFT$ and $INTEGRAL$ 
provide the best opportunity to study the fraction of obscured AGN in the local
universe in the least biased way. Note that for Compton-thick cases (N$_H$ $>$ 
10$^{24}$ cm$^{-2}$), even soft gamma-ray emission could be significantly 
attenuated, thus soft gamma-ray selected samples are not complete to Compton-thick sources.
Claims have been made based on the soft gamma-ray selected samples, that the 
fraction of obscured AGNs in the local universe significantly decrease with 
increasing luminosity.
In this letter we point out that two correction have to be made to the samples
to study of the fraction of obscured quasars. The
corrections are: a) radio loud AGNs have to be excluded since their X-ray 
emission might be dominated by the jet component thus the measured luminosity 
and the obscuration does not reflect the intrinsic values in the nuclei; 
b) Compton thick sources have to be excluded too since their
soft gamma ray emission are also strongly attenuated by Compton scattering
and their intrinsic luminosities are hard to estimate.
After the corrections, we find only marginal decrease in the fraction of
obscured AGN with luminosity in the local universe. Larger samples
are required to reach a more robust conclusion.

\section{The Samples}
\subsection{The $INTEGRAL$ Sample}
Bassani et al. (2006) provides an AGN sample selected in the 20 -- 100 keV
band with the IBIS on $INTEGRAL$. The sample 
contains 62 active galactic nuclei (14 of them are unclassified) above a 
flux limit of $\sim 1.5\times10^{-11}$ ergs cm$^{-2}$ s$^{-1}$. 
B06 listed the available column densities (obtained from archival 
X-ray spectra) for 35 sources with redshifts. 
In this letter we provide an updated list of $N_H$ 
(and references) with more recent measurements
available in literature (mostly from $Chandra$ or $XMM$ observations, Table 1)
\footnote{We note that simply adopting the $N_H$ in B06 does
not significantly affect out main results in this paper.}.
Within this subsample, 10 are radio loud
(including five blazars), five are Compton-thick (N$_{H} \geqslant 10^{24}$ 
cm$^{-2}$), the rest 20 are radio quiet
and Compton-thin. The luminosities in 
the 20 -- 100 keV band were 
calculated by assuming a Crab-like spectrum.

\subsection{The $SWIFT$ Sample}
The first 3 months of the $SWIFT$ Burst Alert Telescope (BAT) high Galactic
latitude survey provide a sample of 14 -- 195 keV band selected sources
(M05) with a flux limit of $\sim$ 10$^{-11}$ ergs
cm$^{-2}$ s$^{-1}$ and $\sim 2\arcmin.7$ (90\% confidence)
positional uncertainties for the faintest sources.
86\% of the 66 high-latitude sources were identified.
Twelve are Galactic-type sources, and 44 are identified with known AGNs.
The luminosities in the 14 -- 195 keV band were
calculated by assuming a typical power-law spectrum ($\Gamma \sim 1.7$).
M05 listed the absorption column density derived from 
literature or archive X-ray spectra for 39 AGNs.
We also provide an updated list of $N_H$ (and references) with more 
recent measurements available in literature (Table 1). The major changes
are that two more Compton-thick sources (NGC 1365 and NGC 7582\footnote{Both
NGC 1365 and NGC 7582 show quick Compton-thick/Compton-thin transitions, see
Risaliti et al. 2005 and Turner et al. 2000.}) are classified.
Within the 39 sources, there are 7 radio loud sources (including three blazars),
4 Compton-thick sources and 28 radio quiet Compton-thin Seyfert galaxies.
\subsection{The Combined Sample}
In order to increase the number of soft Gamma ray selected AGNs, we combine the 
$INTEGRAL$ Sample and the $SWIFT$ Sample. Note that there are 10 objects which
were detected by both of them (see Table 1). Using these 10 objects, we obtained a linear scaling factor from the $SWIFT$ luminosity to the $INTEGRAL$ one (see Fig. 1).
With the scaling factor, we translate all the $SWIFT$ luminosity in 14 -- 195
keV to the the $INTEGRAL$ luminosity in 20 -- 100 keV. The combined 
sample contains 42 radio quiet Compton-thin objects, 14 radio loud sources 
(including 6 blazars) and eight Compton-thick objects.

\section {Results and Discussion}
In Fig. \ref{swift} we plot the X-ray absorption column density versus
the SWIFT luminosity for the 39 AGNs in the SWIFT sample. Consistent with
M05, we find a clear drop of $N_H$ with increasing luminosity. 
Since soft gamma-ray selected samples are not biased to obscuration 
(except for Compton thick sources), the sample selection is identical
for both obscured and unobscured sources. We can simply study the correlation
between $N_H$ and L$_X$ (or the luminosity distribution for obscured/unobscured
sources) in the samples without calculating the luminosity 
function or folding luminosity function model with the
sample selection.
We performed the Spearman Rank (SR) statistic 
to give the correlation between $N_H$ and L$_X$.
We found a significant
anti-correlation between $N_H$ and L$_X$ at 99.99\% level. 
We note that while a flux-limited sample would introduce spurious
correlation between luminosity and redshift (or between luminosities
in different bands), such an effect will not affect our study since
the measurement of $N_H$ is independent of redshift for the local
samples.
With the radio loud and Compton thick sources excluded, the confidence
level of the anti-correlation significantly drops to 98\%.
18 out of 28 radio quiet Compton-thin objects show
$N_H$ $\gtrsim$ 10$^{22}$ cm$^{-2}$, yielding a fraction of 64\%.

In Fig. \ref{integral} we plot $N_H$ versus INTEGRAL luminosity for the
35 AGNs in the INTEGRAL sample. As B06 has claimed,
the fraction of absorbed AGNs in the whole sample significantly decreases 
with the 20 -- 100 keV luminosity. We found an obvious 
anti-correlation between $N_H$ and L$_X$ with a confidence level of 99.7\%.
However, the anti-correlation is obviously dominated by the radio loud and 
Compton-thick objects. Excluding radio loud and Compton-thick objects,
we found no more evidence of anti-correlation (the null hypothesis
gives a probability of 0.92).
14 out of 20 radio quiet Compton-thin objects show
$N_H$ $\gtrsim$ 10$^{22}$ cm$^{-2}$, yielding a fraction of 70\%.

We perform the same statistical analysis to the combined sample described
in \S2 (see Fig. \ref{combine}). Similarly, we find strong anti-correlation
(with confidence level $>$ 99.99\%) between $N_H$ and L$_X$ in the whole sample, but the anti-correlation
is mainly due to the radio loud and Compton-thick objects. With them excluded,
the confidence level of the anti-correlation drops to 94\%.
A KS test also gives similar results that the luminosity distribution of obscured
sources (with $N_H$ $\gtrsim$ 10$^{22}$ cm$^{-2}$)
differs from that of unobscured ones at confidence level of
$>$99.99\% and 95\% respectively before and after excluding
the radio loud and Compton thick sources.
In the combined sample, 29 out of 42 radio quite Compton-thin objects
show $N_H$ $\gtrsim$ 10$^{22}$ cm$^{-2}$, yielding a fraction of 66.7\%.
For 8 luminous objects with L$_{20-100keV}$ $>$ 10$^{44}$ erg s$^{-1}$,
3 show $N_H$ above 10$^{22}$ cm$^{-2}$, yielding a fraction of 38\%.
The drop of the fraction is marginal with a confidence level of 91\%, consistent
with those from the correlation test and KS test.
Note that assuming a powerlaw spectrum with photon index of $\sim$ 2.0, 
L$_{20-100keV}$ is directly comparable to L$_{2-10keV}$.

A recent study by Shinozaki et al. (2006) built a complete flux-limited
sample of bright AGNs from HEAO-1 all-sky catalogs, with a maximum redshift
in the sample of 0.329. Such a sample is ideal to study the population of
obscured quasars in the local universe. Shinozaki et al. found no AGN
in the high luminosity high-intrinsic absorption regime (log L$_X[erg/s]$ 
$>$ 44.5, log $N_H[cm^{-2}]$ $>$ 21.5) in their sample, where $\sim$ 5 AGNs
are expected if assuming a constant fraction of obscured AGNs with luminosity.
Considering that there are 8 AGNs with log L$_X[erg/s]$
$>$ 44.5 and log $N_H[cm^{-2}]$ $<$ 21.5 in their sample, the
difference is at a confidence level of 98\%. We note that 3 out of 8 
sources are obviously radio-loud (3C 273, III Zw2 and PG 1425+267).
Removing the radio-loud sources, the confidence level of the difference
drops to 90\%. This indicates that although the sample presents a trend
of decreasing fraction of obscured AGNs with luminosity, the decrease is
not statistically solid. 

We conclude that the significant decrease with luminosity in the fraction 
of obscured AGNs 
in the local universe might be mainly due to the radio loud and 
Compton-thick objects in the samples. With these objects excluded, we find only
marginal evidence of decrease in the
SWIFT and Shinozaki sample, but no evidence of decrease in the INTEGRAL 
sample which has a similar sample size. 
Much larger samples are required to reach a more robust result on the fraction
of obscured quasars in the local universe.
\acknowledgments
The work was supported by Chinese NSF through NSFC10473009, NSFC10533050 and the CAS "Bai Ren" project at University of Science and Technology of China.

\begin{deluxetable}{lccccc}
\tabletypesize{\scriptsize}
\tablewidth{0pc}
\tablecaption{Soft Gamma-ray selected AGNs with an updated list of $N_H$}
\tablehead{
\colhead{Source} & \colhead{z} & \colhead{Type$^c$} & \colhead{N$_H$} & Ref. &
\colhead{$\log{L}$} 
}
\startdata
Bassani et al. 06 &  INTEGRAL &&&& log L$_{20-100keV}$ \\
\hline 
QSO B0241+62 & 0.044 & RQ & 22.2 & 1 & 44.48  \\
......\\
\hline
Markwardt et al. 2005 &  SWIFT &&&& log L$_{14-195keV}$-0.018\\
\hline
Mrk 348 & 0.0150 & RQ & 23.1 & 14 & 43.68  \\
......\\
\enddata\\
Note. -- The complete version of this table is in the electronic edition of
the Journal.  The printed edition contains only a sample.  $^a$ Sources included by both Bassani et al. 2006 and Markwardt et al. 2005.  $^b$ QSO 0836+710.  $^c$ RQ = radio quiet; RL = radio loud; Bl = blazar; Co = Compton thick.  References.--(1) B06; (2) Matt et al. 2006; (3) Dewangan, Griffiths \& Schurch 2003; (4) Beckmann et al. 2005; (5) Beckmann et al. 2004; (6) M05; (7) Matt et al. 2004; (8) Steenbrugge et al. 2003; (9) Steenbrugge et al. 2005; (10) Matsumoto et al. 2004; (11) Lewis et al. 2005; (12) Bianchi et al. 2005; (13) Foschini et al. 2006; (14) Akylas et al. 2002; (15) Risaliti et al. 2005; (16) Ballantyne et al. 2004; (17) Vaughan et al. 2004; (18) Colber et al. 2005; (19) Jiang, Wang \& Wang 2006; (20) Yaqoob et al. 2005; (21) Costantini et al. 2005; (22) Gibson et al. 2005; (23) Turner et al. 2000; (24) Bianchi et al. 2003
\end{deluxetable}

\begin{figure}
\plotone{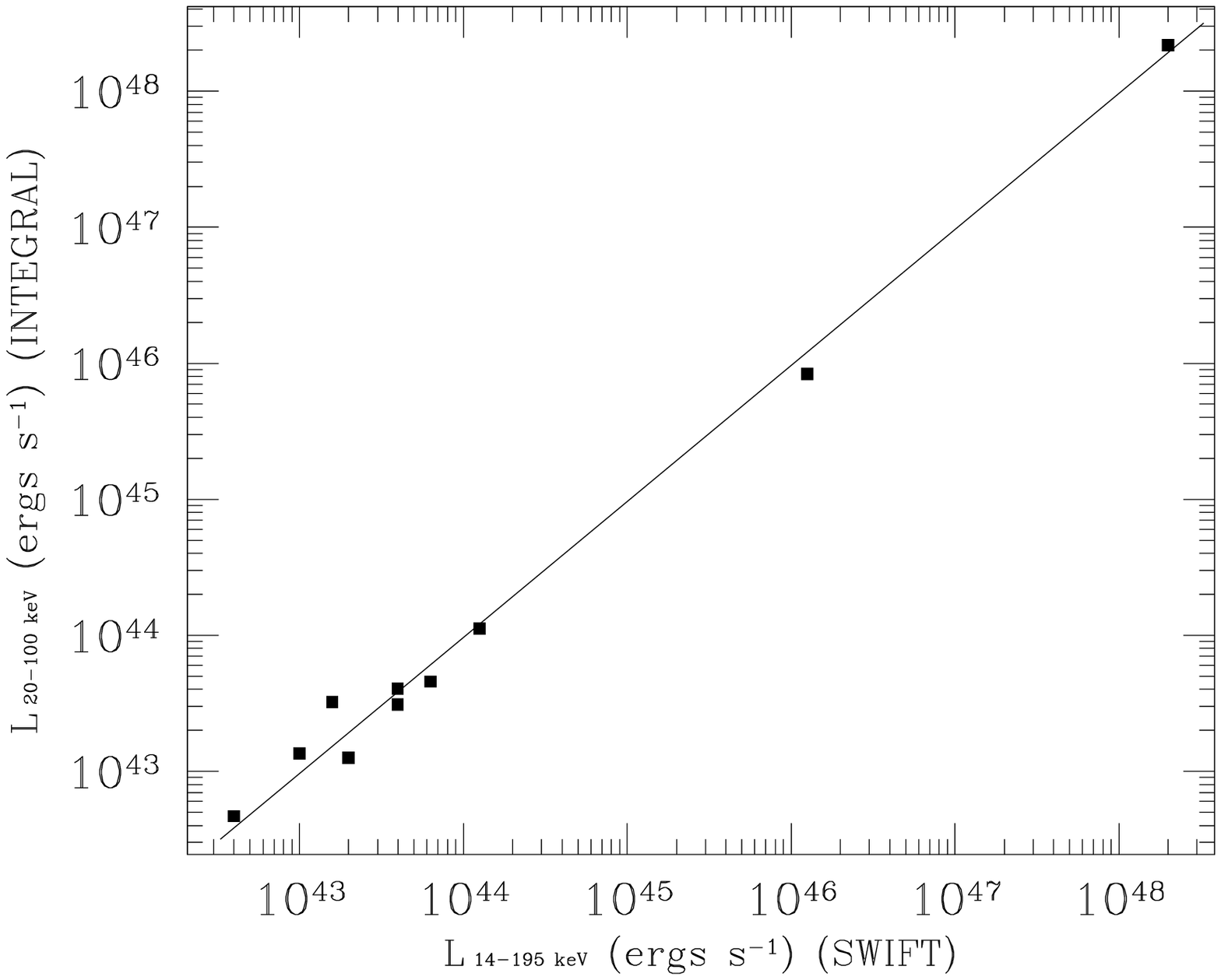}
\caption{
The INTEGRAL luminosity in the 20 -- 100 keV band (from Bassani et al. 2006) 
versus the $SWIFT$ luminosity in the 14 -- 195 keV band (from Markwardt et al. 
2005) for 10 AGNs which were detected by both instruments.
The solid line shows a linear correlation between two luminosities:
Log (L$_{20-100keV}$) = Log (L$_{14--195keV}$) - 0.018.
}
\label{scaling}
\end{figure}

\begin{figure}
\plotone{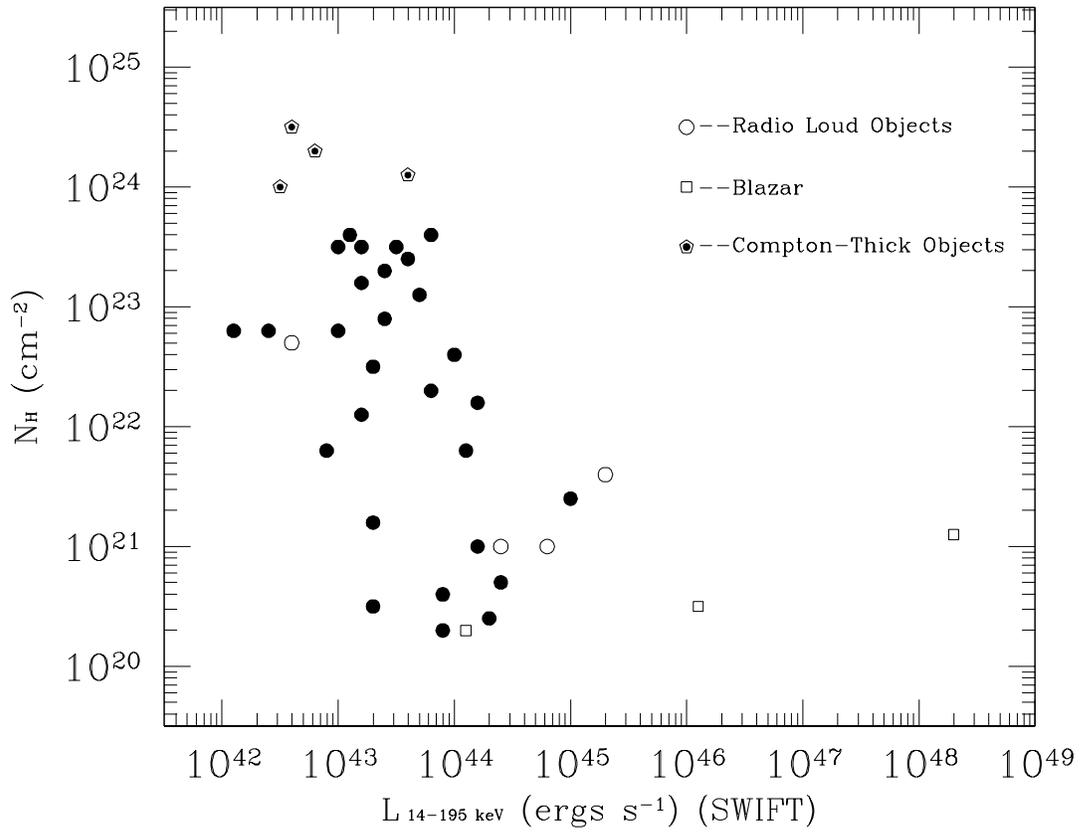}
\caption{
The X-ray absorption column density $N_H$ versus the $SWIFT$ luminosity in the 14 -- 195 keV band for the 39 AGNs in Markwardt et al. (2005).
}
\label{swift}
\end{figure}

\begin{figure}
\plotone{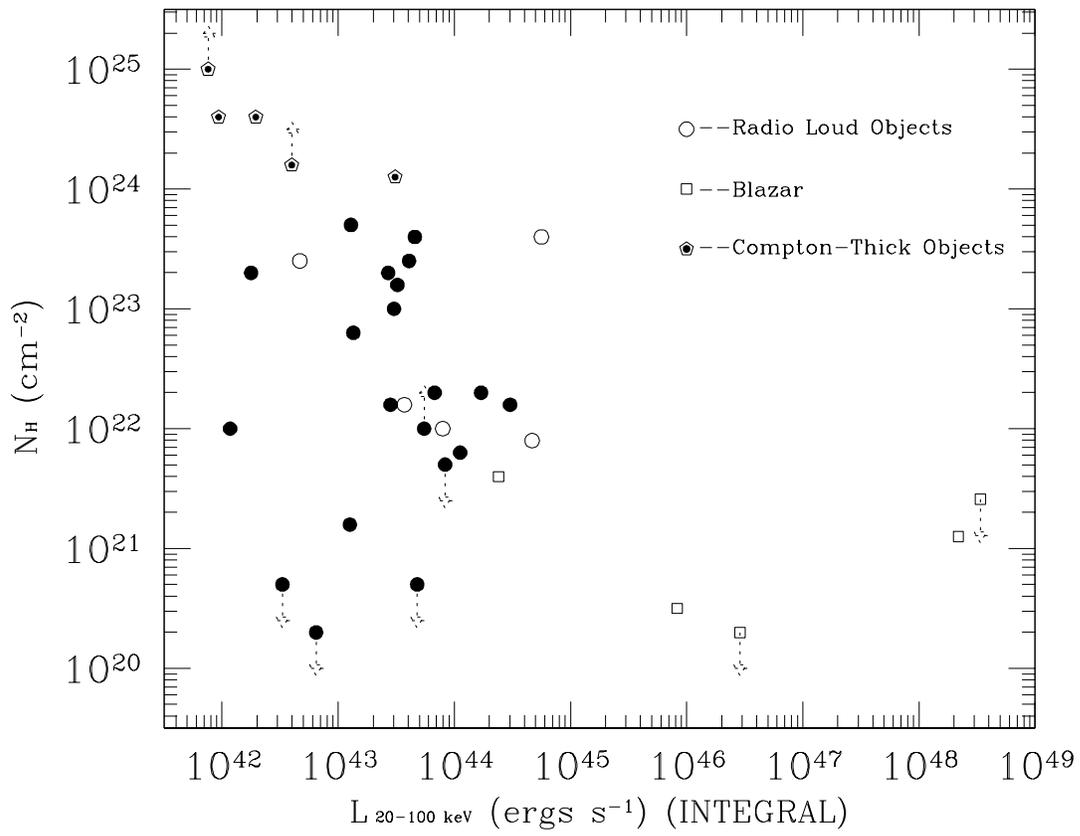}
\caption{
The X-ray absorption column density $N_H$ versus the INTEGRAL luminosity in the 20 -- 100 keV band for the 35 AGNs in Bassani et al. (2006).
}
\label{integral}
\end{figure}

\begin{figure}
\plotone{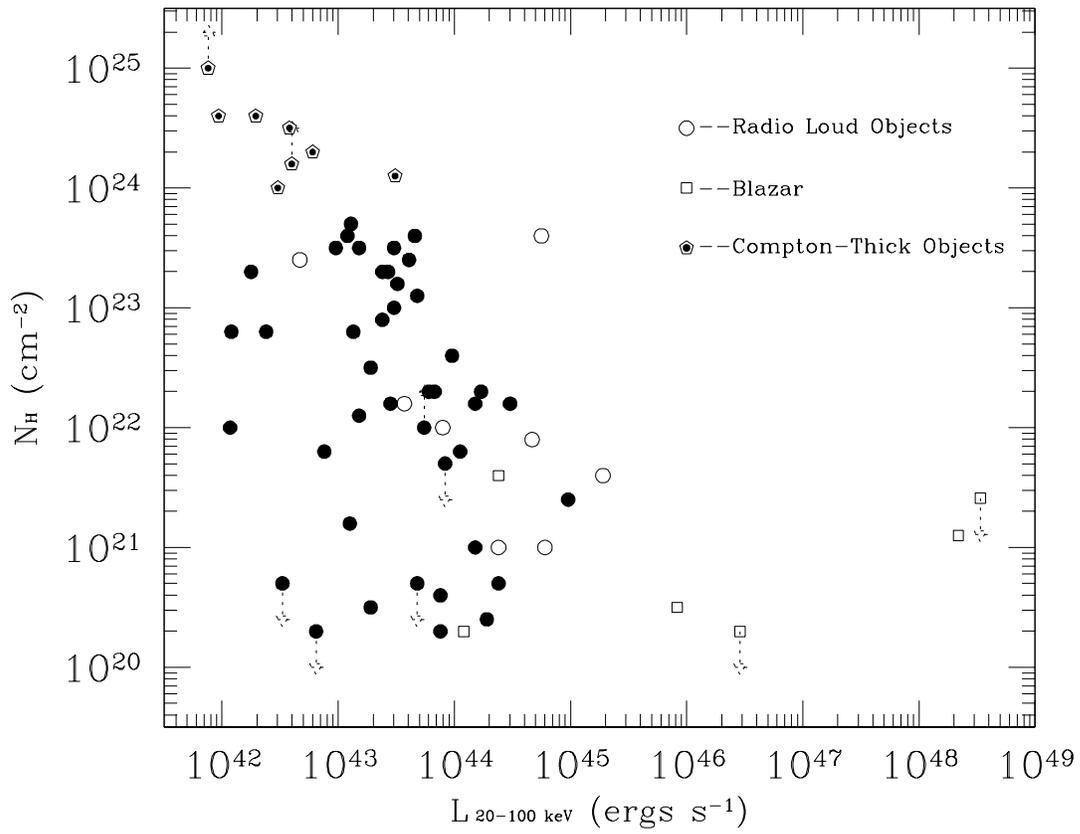}
\caption{
The X-ray absorption column density $N_H$ versus the INTEGRAL luminosity 
for the 65 AGNs in the combined sample.
}
\label{combine}
\end{figure}

\end{document}